\documentclass[aps,prb,groupedaddress,showpacs]{revtex4}
\usepackage{amsmath}
\usepackage{amssymb}
\usepackage{graphicx}

\begin{document}


\title{Phononless thermally activated transport through a disordered array of quantum wires}

\author{A. L. Chudnovskiy}
\affiliation{I. Institut f\"ur Theoretische Physik, Universit\"at Hamburg,
Jungiusstr. 9, 20355 Hamburg, Germany}


\date{\today}

\begin{abstract}
Phononless plasmon assisted transport through a
long disordered array of finite length quantum wires is
investigated analytically. Two temperature regimes, the low- and the
high-temperature ones, with qualitatively different temperature dependencies of
thermally activated resistance are identified.  The characteristics of plasmon assisted and
phonon assisted transport mechanisms are compared.
Generically strong electron-electron interaction
in quantum wires results in  a qualitative change of the temperature dependence
of thermally activated resistance in comparison to phonon assisted transport.
At high temperatures, the thermally activated resistance is determined by the
Luttinger liquid interaction parameter of the wires.
\end{abstract}

\pacs{73.63.-b,72.10.Di,73.23.-b}

\maketitle


\section{Introduction}

It is a well established fact that the single-electron transport through a one-dimensional system with random scattering is suppressed at zero temperature and infinitesimally small applied voltage \cite{Berezinskii}. The physical mechanism of the suppression of transport is known to be the Anderson localization phenomenon, which is explained as a constructive interference of initial and back-scattered waves that enhances the return probability after scattering by an impurity potential. Eventually the constructive interference  leads to the localization of the single particle wave  function in a one-dimensional disordered system. A natural conclusion follows, that suppression of the coherence of the single particle propagation leads to delocalization and favors transport through the system. Indeed, at finite temperature electrons couple to thermally activated bosonic excitations in the environment (usually phonons),  which results in the decoherence of the electron motion and in the thermally activated electron transport \cite{Efros}. 

A basic theoretical question remains, whether the thermally activated transport can be induced by electron-electron interactions alone, that is without an external bath of bosonic excitations like phonons. The mechanism of a phononless thermally activated transport consists of dephasing of the electron wave function by electron-electron scattering that results from e-e interactions. In the case of decoherence by interactions, the role of bosonic bath is played by the electron-hole pairs, or charge fluctuations, that are excited thermally or in the result of electron-electron scattering. A microscopic mechanism for creating a bosonic bath out of localized electrons has been first proposed in Ref. \cite{Tigran}.

Recently the problem of thermally activated transport in a disordered one-dimensional wire with interactions has been addressed in several papers \cite{Gornyi,Malinin,Fogler03,Aleiner,Mirlin}. In the absence of disorder, a one dimensional wire with interactions is described by the Tomonaga-Luttinger liquid model. This model allows exact treatment of the low-energy physics in the interacting system using the bosonization technique \cite{Haldane,JvD}. At the same time it is known that the Luttinger liquid is destroyed virtually by any small amount of random scattering \cite{Kane-Fisher}. Therefore, in the localized regime that is dominated by disorder the Luttinger liquid description is inapplicable.

The two abovementioned limits suggest two approaches to the problem of thermally activated transport. The first one is based on the description of  one dimensional wire as a disordered Fermi liquid.
In that approach, the interactions are treated perturbatively. In recent works  a  metal-insulator transition at finite temperature has been reported in a disordered weakly interacting systems.
The mechanism of transport at finite temperature consists in formation of a many particle delocalized state in which correlations between distant localized  single particle states are established through thermally excited multiple particle-hole pairs \cite{Aleiner,Mirlin}.

The other approach treats the localized system as a pinned charge density wave. The thermally activated transport is described as a propagation of instantons through the system \cite{Malinin,Fogler03}. The theoretical description has then much in common with bosonization. It is applicable for not too weak interactions.  That description suggests a thermally activated behavior of conductance similar to the variable range hopping. Most importantly,  an external bath of bosonic excitations with continuous spectrum is necessary to facilitate the transport \cite{Malinin}.

The electron-electron scattering can be equivalently reformulated as a scattering of electrons by charge fluctuations (plasmons). If the effective interaction  is short-range, the scattering rate by electron-plasmon collisions is given by the Fermi golden rule. The application of the Fermi golden rule is based on the assumptions of complete loss of coherence of plasmons between two subsequent collision events. This is a good approximation in a system with strongly screened interactions, which have a Fermi liquid ground state, such as a three dimensional metall. It becomes progressively worse with lowering the dimensionality of the system. An extreme example of a system with highly coherent plasmons is given by a pure one-dimensional system, or quantum wire. In that case, the charge density fluctuations are well defined quasi-particles with an infinite coherence time, and theoretically such system is described by a Tomonaga--Luttinger (TL) liquid model \cite{Haldane,JvD}. At the same time, the electron itself ceases to be a quasiparticle that propagates coherently through a   Tomonaga--Luttinger liquid. By entering the TL liquid form outside, an electron decays into charge and spin excitations. This decay process has a very slow power law character in time, which is also known as a demonstration of the Anderson orthogonality catastrophe \cite{Anderson}. For such systems a straightforward application of the Fermi golden rule leads to incorrect results as it
will be shown on an example considered in the present paper.

\begin{figure}
\includegraphics[width=12cm,height=8cm,angle=0]{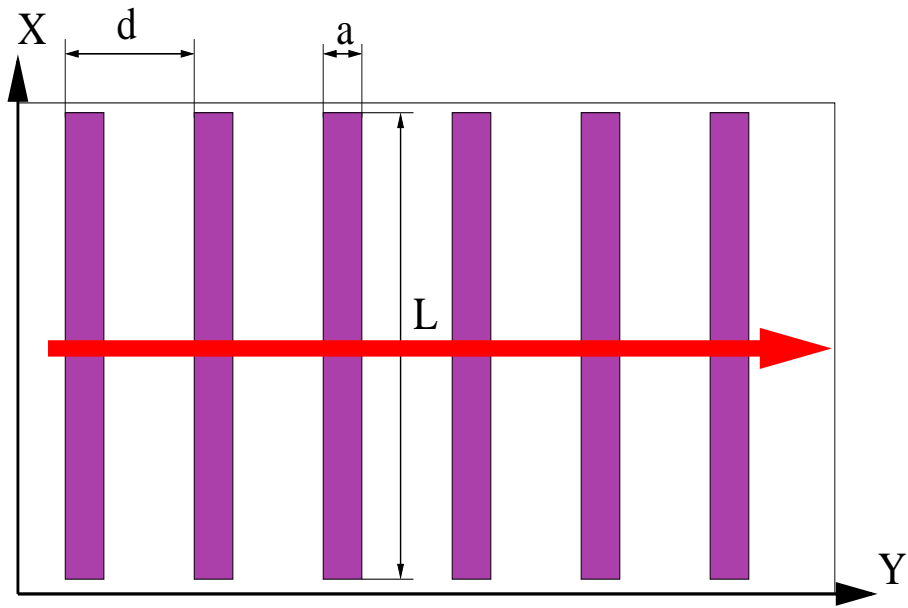}%
\caption{Geometry of the model. The arrow shows the direction of the current.  \label{figsetup}}
\end{figure}

In this paper we investigate phononless thermally activated transport through a quasi-one-dimensional system formed by a parallel arrangement of conducting wires (see Fig. \ref{figsetup}). Each wire has a finite length $L$ and the transport direction is perpendicular to the wires. We identify the regime, where the localization length of plasmons in the array exceeds very much the single particle localization length. In that regime we show that charge-density fluctuations (plasmons) in the array can act as the agent promoting thermally activated transport, thus providing the possibility for phononless inelastic transport.
The plasmon assisted transport mechanism considered in the present paper is essentially the same as in
Refs. \cite{Aleiner,Mirlin}. However, the adopted model exhibits a crossover from a Fermi liquid metal (or Anderson insulator in the presence of disorder) at zero temperature to a strongly interacting sliding Luttinger liquid system at higher temperatures. Therefore,  perturbative treatment of interactions at finite temperature is inapplicable to the system under consideration.
As the result of generically strong plasmon-electron coupling in a quantum wire,
the features of plasmon and phonon assisted transport are qualitatively  different.
We provide a qualitative explanation of plasmon assisted transport, identify the transport regimes, where the features of plasmon and phonon assisted transport are either similar or substantially different, and
derive analytic expressions for the temperature dependence of the thermally activated
resistance.

Hopping transport in the considered model is of much relevance to a number of experimental setups,
including quantum wire arrays in heterojunctions \cite{Mani}, carbon nanotube films \cite{deHeer},
atomic wires on silicon surface \cite{Himpsel}, and stripe phases \cite{Fogler}.
At finite length of constituent wires, such systems represent particular
examples of granular arrays, where a one-dimensional wire  plays the role of a grain.
Considered as a granular array, the array of parallel quantum wires is rather peculiar
because of the very long charge relaxation time in a one-dimensional wire.
Due to this peculiarity, the theoretical description of thermally activated transport
in arrays of long quantum wires requires  taking into account the charge
dynamics and treatment of interactions beyond the capacitive model adopted in recent
theoretical investigations of transport through disordered granular arrays
\cite{AGK,Efetov,Fogler03}.

The rest of the paper is organized as follows. In Sec. \ref{sec-model} the theoretical
model is introduced in detail and the relevant physical regime is identified. In Sec.
\ref{sec-RR} we derive general equations for the thermally activated resistance of the
array. Results of detailed calculations in the low- and high-temperature regimes are
presented in Secs. \ref{sec-lowT} and \ref{sec-highT} respectively. In Sec. \ref{sec-phonon}
the calculation of the thermally activated phonon assisted transport for the considered model
is presented in order to make a comparison with the plasmon activated transport. Our
findings are summarized in Sec. \ref{sec-concl}.

\section{Theoretical model}
\label{sec-model}

The model we formulate below is special, because it combines two seemingly incompatible
features: {\bf i)} it is strongly disordered for single electron transport;
{\bf ii)} it is much weaker disordered for propagation of plasmons.

Consider a one dimensional array of parallel  identical quantum wires of length
$L$ and diameter $a$ placed regularly with the interwire distance $d$, $L\gg d\gg a$.
We investigate transport in the
direction perpendicular to the wires (see Fig. 1). The spectrum of low-energy plasmons in a single 
isolated wire is equidistant with energies
\begin{equation}
E_{i,n}=\frac{\pi u_i}{L}n,
\label{E_in}
\end{equation}
where $L$ is the length of the wire, and $u_i$ is the plasmon
velocity along the wire $i$.
For identical wires at regular positions, the intra- and
interwire interactions between the charge density fluctuations do not change along the array.
Then each localized plasmon level broadens into a plasmon band
with truly continuous spectrum, quite analogously to the formation of electronic bands
in the tight binding model. The role of the hopping in the tight-binding model
for plasmons is played by the matrix element of the charge-density interactions between the
neighbor wires.
The formation of plasmon bands is reflected by the dependence of
the plasmon velocity along each wire on the wave vector $p$ along the array (that is,
perpendicular to the wires)
$u_i\rightarrow u(p)$. The particular form of plasmon dispersion depends on details of the
interwire interactions, yet the function $u(p)$ should be periodic with
a period of one Brillouin zone. That is why we choose the specific form
\begin{equation}
u_p=v_0+v_1\cos(\pi p), \ -1<p\leq 1.
\label{u_p}
\end{equation}
The chosen form of dispersion can be considered as the first two terms of the Fourier
expansion of some general dispersion law. The plasmon energy within a band centered around
the level $n$ is given by $\epsilon_{n}(p)=\frac{\pi u_p}{L}n$.

\subsection{Single particle localization length}

Since we are interested in the
single particle transport in the direction perpendicular to the wires, we consider
in this section only the transverse components of single particle  wave functions,
assuming the factorization of the wave function into the transverse and longitudinal
(with respect to  the direction of the wires) parts.

Suppose the typical height of
potential barriers between the neighbor wires is very large. Let us describe the
array without interactions by a tight-binding model.
Then a single particle wave function has most of its weight inside the
wires. We approximate a Vannier wave function localized in a single wire with number $n$
by the form
\begin{equation}
\psi_n(y)=\psi_0\left\{\theta\left(y+\frac{a}{2}-nd\right)+
\theta\left(nd+\frac{a}{2}-y\right)+e^{-\lambda |nd-y|}\left[\theta\left(n d-\frac{a}{2}-y\right)+
\theta\left(y-\frac{a}{2}-nd\right)\right]\right\},
\label{Vannier-wf}
\end{equation}
where $\theta(y)$ is a step function.
Here and everywhere further $y$ denotes the coordinate in the
direction perpendicular to the wires, that is in the transport direction.
Using the normalization condition for $\psi_n(y)$, the value of $|\psi_0|^2$ is found to be
\begin{equation}
|\psi_0|^2=\frac{\lambda}{a\lambda+1}.
\label{norm}
\end{equation}
In the absence of disorder and interactions  the single particle motion is described by
very narrow energy bands with dispersion $\epsilon_k=t\cos(kd)$,
$(-\frac{\pi}{d}<k\leq\frac{\pi}{d})$.
With the help of (\ref{Vannier-wf}) the bandwidth $t$ that equals the
hopping energy in the equivalent tight-binding model can be estimated as
\begin{equation}
t\approx \frac{H\lambda d}{a\lambda+1} e^{-\lambda d},
\label{ev-t}
\end{equation}
where
$\lambda$ describes  the decay of the single particle wave function inside the barrier of
the height $H$,
\begin{equation}
\lambda\approx\sqrt{\frac{2m}{\hbar^2}H}.
\label{ev-lambda}
\end{equation}

Now let us introduce the disorder as a random height of the energy barrier between the neighbor
wires. Furthermore, the charge transfer between the wires is accompanied
by Coulomb blockade effects. Due to the random environment (local concentration of charged
impurities) around each wire, the charging energy $E_c$ is random, which represents another
source of randomness in the model. Such disorder induces fluctuations of the decay parameter $\lambda$ thus rendering $t$
random. As the result, the single particle wave functions become localized. In 1D the single
particle localization length $\xi_1$  is of the order of the mean free path $l_f=v_F \tau_f$.
The mean free time $\tau_f$  is in turn related to the
fluctuations of the heights of the tunneling barriers
\begin{equation}
\langle\delta H(n) \delta H(n')\rangle=\frac{1}{2\pi\nu_1 \tau_f d}\delta_{n,n'}.
\label{deltaH}
\end{equation}
Here $n$ and $n'$ denote the numbers of the wires adjacent from the left to the barrier,
and we assigned a length scale $1/d$ to the $\delta$-function.
Furthermore, $\nu_1$ is a single particle density of states, which is given by $\nu_1=\frac{1}{td}$
in the center of the band. Now let us establish a connection between the mean free path $l_f$
and the fluctuations $\delta\lambda$ of the decay parameter of the wave function.
Using the relation (\ref{ev-lambda}) we obtain
\begin{equation}
\delta{\lambda}=\sqrt{\frac{m}{2\hbar^2\overline{H}}}\delta H.
\label{ev-dlambda}
\end{equation}
Further we solve (\ref{deltaH}) for $\tau_f$ and express $\delta H$ through $\delta\lambda$
with the help of (\ref{ev-dlambda}). In the result we obtain
\begin{equation}
l_f=\frac{\overline{\lambda}^4d^3e^{-2\overline{\lambda} d}}{8\pi
\langle{\delta\lambda}^2\rangle(a\overline{\lambda}+1)}.
\label{ev-lf}
\end{equation}
Imposing the condition for the single particle to be localized within two nearest
neighbor wires $l_f=d$, we obtain for $\langle{\delta\lambda}^2\rangle$
\begin{equation}
\langle{\delta\lambda}^2\rangle=\frac{\overline{\lambda}^4d^2e^{-2\overline{\lambda}d}}{8\pi
(a\lambda+1)}.
\label{lf=d}
\end{equation}

\subsection{Plasmon localization length}

The random height of interwire tunneling barriers affects the interwire charge density
interactions  through the randomness of the
transverse part of the wave functions $\psi_i$. The matrix element of the charge density
interaction between the neighbor wires with numbers $1$ and $2$
(direct part of Coulomb interaction) is calculated as
\begin{eqnarray}
\nonumber &&
V_{12}\propto\int_{-\infty}^{\infty}dy_1dy_2\frac{|\psi_1(y_1)|^2
|\psi_2(y_2)|^2}{|d+y_2-y_1|} \\
\nonumber &&
=\int_{-a/2}^{a/2}dy_1dy_2\frac{|\psi_1(y_1)|^2
|\psi_2(y_2)|^2}{d+y_2-y_1}+
\left(\int_{-\infty}^{-a/2}+\int_{a/2}^{\infty}\right)dy_1\int_{-a/2}^{a/2}dy_2
\frac{e^{-2\lambda |y_1|}|\psi_2(y_2)|^2}{|d+y_2-y_1|}+ \\
&&
\int_{-a/2}^{a/2}dy_1\left(\int_{-\infty}^{-a/2}+\int_{a/2}^{\infty}\right)dy_2
\frac{e^{-2\lambda |y_2|}|\psi_1(y_1)|^2}{|d+y_2-y_1|}+
\left(\int_{-\infty}^{-a/2}+\int_{a/2}^{\infty}\right)dy_1dy_2
\frac{e^{-2\lambda (|y_1|+|y_2|})}{|d+y_2-y_1|}.
\label{directClmb}
\end{eqnarray}
We took into account that the wave function decays exponentially inside the barrier (for $|y_i|>a/2$).
For $\lambda d\gg 1$ the main contribution to the matrix element is given by the first
term in (\ref{directClmb}) that relates to the interaction of charge densities inside the
wires. That term results in
\begin{equation}
V_{i,i+1}\sim a^2|\psi_i|^2|\psi_{i+1}|^2/d,
\label{Vi,i+1}
\end{equation}
where
\begin{equation}
|\psi_i|^2=\frac{1}{a}\int_{-a/2}^{a/2}dy|\psi_i(y)|^2.
\label{psi-av}
\end{equation}
Note that being integrated over the wire crosssection, the value  $|\psi_i|^2$ in insensitive
to random variations of the wave function across the wire. The contributions from other terms
in (\ref{directClmb}) are suppressed as $1/\lambda$. The effect of  the random
potential barrier on the quantity $|\psi_i|^2$  is encoded in the decay factor $\lambda_i$.
Setting in (\ref{norm}) $\lambda_i=\overline{\lambda}+\delta\lambda_i$ we obtain the
fluctuation of $|\psi_i|^2$ in the form
\begin{equation}
\delta|\psi_i|^2=\frac{\delta\lambda_i}{(a\overline{\lambda}+1)^2}.
\label{deltapsi}
\end{equation}
Furthermore, using (\ref{Vi,i+1}) we estimate the fluctuation of the matrix element of the
interaction as
\begin{equation}
\delta V_{i,i+1}\sim \frac{2\overline{\lambda}\delta\lambda_i a^2}{d(a
\overline{\lambda}+1)^3}.
\label{deltaV}
\end{equation}

In the bosonized form, the density-density interaction between the neighbor wires reads
\begin{equation}
H_{i,i+1}=\int_0^L dx  V_{i,i+1}\left(\partial_x\phi_i(x)\right)\left(\partial_x\phi_{i+1}(x)\right).
\label{Hi,i+1}
\end{equation}
Using the mode expansion of the bosonic fields $\phi_i(x)$, $\phi_{i+1}(x)$ \cite{JvD}
\begin{equation}
\phi_i(x)=-\sum_q\frac{1}{n_q}\left(e^{-iqx}\hat{b}_{i,q}+e^{iqx}\hat{b}^{\dagger}_{i,q}\right)
\label{modeexp}
\end{equation}
together with the relations $q=\frac{\pi}{L}n_q$, $\omega_q=uq$, where $u$ is the phase velocity of
a plasmon along the wire, we rewrite (\ref{Hi,i+1}) in the form of a hopping Hamiltonian for the bosons represented by operators $\hat{b}_{i,q},\hat{b}^{\dagger}_{i,q}$
\begin{equation}
H_{i,i+1}=V_{i,i+1}\sum_q\frac{2\pi\omega_q}{u}\left(\hat{b}^{\dagger}_{i,q}\hat{b}_{i+1,q}+
\hat{b}^{\dagger}_{i+1,q}\hat{b}_{i,q}\right).
\label{i,i+1hop}
\end{equation}

In the result of random interwire interactions,
the plasmons become localized in the direction along the array.
The plasmon localization length can be evaluated as $\xi_p=u_{gr}\tau^{\omega}_p$, where
$\tau^{\omega}_p$ denotes the mean free time of a plasmon mode with frequency $\omega$, and $u_{gr}$ is the group velocity of plasmons. The plasmon mean free time is related to the fluctuations of the
matrix element of the interwire charge density interactions
\begin{equation}
\left(\frac{2\pi\omega_q}{u}\right)^2\langle(\delta V_{i,i+1})^2\rangle=\frac{1}{2\pi d \nu_p\tau^{\omega}_p},
\label{def-taup}
\end{equation}
where $\nu_p=\frac{L}{\pi^2v_1 n}$ is the plasmon density of states in the middle of the band
$n$.
In turn, the velocity $v_1$ is related to the average value of the charge density interactions
$\overline{V}=\langle V_{i,i+1}\rangle$ that in turn characterizes the width of plasmon bands in the
pure system.  From the
dispersion law of the lowest plasmon band $\epsilon_p=\frac{\pi}{L}(v_0+v_1\cos(\pi p))$ we infer
\begin{equation}
\overline{V}=\frac{\pi}{L}v_1.
\label{v_1(V)}
\end{equation}
Furthermore, the plasmon group velocity is given by
\begin{equation}
u_{gr}=\frac{1}{d}\frac{d\epsilon_p}{dp}=\frac{\pi^2 d}{L}v_1.
\label{u_gr}
\end{equation}
Using  (\ref{Vi,i+1}) and  (\ref{norm})  the average interaction strength
$\overline{V}$ can be expressed as
\begin{equation}
\overline{V}\sim \frac{\lambda^2 a^2}{d(a\lambda+1)^2},
\label{av-V}
\end{equation}
where $\langle\delta\lambda^2\rangle\ll\lambda^2$ is assumed.
Solving (\ref{def-taup}) for $\tau^{\omega}_p$ and expressing the plasmon density of states for the
first plasmon band $n=1$ through the average interaction strength $\overline{V}$ with the
help of (\ref{v_1(V)}) and (\ref{av-V}), we  obtain the expression for the plasmon mean free time in the form
\begin{equation}
\tau^{\omega}_p\approx\left(\frac{u}{2\pi\omega}\right)^2\frac{d (a\overline{\lambda}+1)^4}{8\langle\delta\lambda^2\rangle a^2}.
\label{tau_p}
\end{equation}
Now let us evaluate the ratio between the plasmon localization length $l_p$ and the
electron localization length $d$. The condition of strong electron localization on the
length $d$ determines the value of $\langle\delta\lambda^2\rangle$ as given by  Eq.
(\ref{lf=d}). Substituting that value of $\langle\delta\lambda^2\rangle$ into the
expression (\ref{tau_p}) and multiplying by $u_{gr}/d$ with $u_{gr}$ given by (\ref{u_gr})
we finally obtain
\begin{equation}
\frac{l_p}{d}\approx \frac{s^2(a\overline{\lambda}+1)^4}{4\omega^2\overline{\lambda}^2d^2}
e^{2\overline{\lambda}d}.
\label{lp/d}
\end{equation}
This result agrees qualitatively with the evaluation of plasmon correlation length in a randomly inhomogeneous Luttinger liquid by Gramada and Raikh \cite{Gramada-Raikh}.
Eq. (\ref{lp/d}) shows that the condition $l_p/d\gg 1$ can be formulated as
\begin{equation}
\overline{\lambda}d\gg\ln\left[\frac{2\omega\overline{\lambda}d}{s(a\overline{\lambda}+1)^2}
\right].
\label{l/d>>1}
\end{equation}
Eq. (\ref{l/d>>1}) defines the regime, where the plasmon localization length exceeds
very much the electron localization length. If, additionally, $l_p$ exceeds the length
of the array in the transport direction (orthogonal to the wires), the plasmon bands
can be applied to describe the plasmon spectrum in the array.

\section{Resistance of a disordered array}
\label{sec-RR}

The resistance of a long disordered one-dimensional array is determined by so-called
breaks, the junctions between two neighboring  wires with exponentially high resistance
\cite{RR}. Let us denote the energy cost to transfer an electron over the break as  $E_a$.
To facilitate the transport over the break, the energy $E_a$ should be
borrowed by absorption of a bosonic excitation.
For two isolated wires forming the break, the matching condition between the energy of a plasmon
mode $n$ in the wire $i$ given by (\ref{E_in}) and the energy $E_a$,
$E_{i,n}=E_a$,  cannot be satisfied for arbitrary $E_a$ because of discreteness of $E_{i,n}$.
However, due to the charge-density interwire interaction (\ref{i,i+1hop}) the energy can
be transferred between excitations localized in different wires. In the regime discussed in previous section the plasmon localization length exceeds the length of the array. Then the plasmon band description of the plasmon spectrum is applicable. Treating the interwire charge density interaction perturbatively, we can write the
transition rate with the absorption of a plasmon similarly to a transition with the absorption
of a phonon using the Fermi golden rule \cite{Efros},
\begin{equation}
\gamma\propto \int dp \sum_{n}\sum_{m,k=0}^{\infty}|V_n(p)|N_B(\epsilon_{n}(p))
f\left(-E_m\right)
\left[1-f\left(E_a+E_k\right)
\right]\delta\left(\epsilon_{n}(p)-E_a-E_m-E_k\right).
\label{Golden-Rule}
\end{equation}
Here $V_n(p)$ is the strength of  interwire charge density interaction for the plasmon mode
$n$, $N_B(\epsilon_{n}(p))$ is the occupation number of the  plasmon mode,
$f(E_m)$ denotes the Fermi distribution and describes the occupation of the
$m$-th single-particle energy level in the wire, $E_m=\frac{\pi v_0}{L}m$.
For narrow plasmon energy bands, the perturbative approach suggests that
when  the energy $E_a$ lies in the gap between the plasmon bands the
hopping over the break is blocked.  This suggestion turns out to be wrong because
of a  conceptual difference between the plasmon and phonon transport mechanisms. Whereas the
phonons represent a bath of bosonic excitations that is independent of electrons,
the plasmons are ``made'' of electrons themselves.
Consequently, while the electron-phonon interaction can generally be
treated perturbatively, the perturbative treatment of plasmons is possible only under special
conditions. The applicability of the perturbative treatment of plasmons is determined by the
relation of two time scales: the characteristic time of plasmon dynamics $t_p$ and the
characteristic time of a single electronic hop $t_h$.
If the Coulomb interaction in a grain is well-screened or the plasmons are strongly
localized, then $t_p$ is the characteristic relaxation time of a plasmon excitation
within a single grain. For  $t_p\ll t_h$ the plasmons can be neglected
in transport. The description of interactions thus reduces to the capacitive model
\cite{AGK,Efetov,Fogler03}.
For the delocalized undamped plasmons, the time $t_p$ is
associated with the formation of an extended in space plasmonic excitation. In that case,
$t_p\ll t_h$ correspond to the regime of a strongly nonlinear coupling between plasmons
and electrons, and the perturbative treatment of plasmons is incorrect.
Plasmons in one-dimensional wires represent a profound example for that regime. In particular,
the relation $t_p\ll t_h$ is always fulfilled at the break.
For the model considered in this paper $t_p\sim L/v_1$. As we show below, due to the strong
electron-plasmon coupling, the nonlinear
effects lead to the creation of plasmon complexes with energies covering  the
whole spectrum continuously, even though the plasmon bands initially are very narrow.
This in turn leads to plasmon assisted transport with a temperature dependence that is
qualitatively different from the case of phonon assisted transport.
In the regime  $t_p\gg t_h$, the effective
interaction time is limited by $t_h$. Then the plasmon dynamics is essentially
independent of the electron dynamics and does not affect the electronic transport.

The resistance of the array is calculated along the lines of
Ref. \cite{RR}.
Let us parameterize the tunneling matrix element between two wires in the form
$
t_{i,i+1}=\exp(-|y_{i,i+1}|/d)
$.
The parameter $y$ can be associated with an effective distance between the two wires.
This effective distance is random, its distribution follows from the distribution of the
heights of potential barriers.
Since a break, being a junction with exponentially large resistance, is not
shorted by other resistances connected in parallel, we can write the resistance of a break in
the form
\begin{equation}
R_{1}=R_0\exp[2|y_{i,i+1}|/d+f(E_a,T)].
\label{R1}
\end{equation}
Here $E_a$ denotes an addition energy to transfer an electron over the break.
We remind that the disorder enters the model only as a random distribution of addition
energies $E_a$ and effective distances $y_{i,i+1}$.
The function $f(E_a,T)$ accounts for the effect of thermally activated
plasmons in the resistance of the break.
According to Ref. \cite{RR},
the probability density $\rho(u)$ for the resistance  $R/R_0=e^u$ is proportional to
$e^{-gA}$, where $A$ is the area in the $(y,E_a)$ phase
space that results in the resistance $e^u$, and $g$ is the linear
density of localized one particle states.
The area $A$ can be calculated as
\begin{equation}
A(u)=a\int_0^{u-f(E,T)=0}dE[u-f(E,T)].
\label{area}
\end{equation}
Then the resistance is given by
\begin{equation}
R=R_0l_y\int_0^\infty du \  e^{u-gA(u)},
\label{R_l}
\end{equation}
where $l_y$ is the length of the array. The function $e^{u-gA(u)}$ is usually strongly peaked, which  justifies  the evaluation of (\ref{R_l}) by the saddle point method. The saddle point value of $u$  corresponds to the resistance of an {\it optimal} break in the chain \cite{RR}.
Therefore, in order to calculate the resistance of the array in the localized regime,
we have to obtain an expression for the resistance of the break $R_1$. Since the break is not
shorted by other resistances, we conclude $R_1=1/\sigma_1$, where $\sigma_1$ is the conductance
of a break.
Assume that the break is formed by a junction between the wires with numbers
$0$ and $1$. We take the position of a pinhole connecting the
two wires as $x$.
In the linear response approximation the current through the break $I_1$ is determined
by  the correlation function \cite{Mahan}
\begin{equation}
X(\tau)=|t_{01}|^2\left\langle T_{\tau}\left(\Psi_0(x,\tau)\Psi^{\dagger}_1(x,\tau)\Psi_1(x,0)
\Psi^{\dagger}_0(x,0)\right)
\right\rangle
\label{X}
\end{equation}
that characterizes the probability of a single hop over the break.
Here $t_{01}$ is the tunneling matrix
element, $\tau$ is an imaginary time. In what follows we  denote
$X_+(\tau)=X(\tau>0)$ and $X_-(\tau)=X(\tau<0)$.

A unique feature of the chosen model is the applicability of the bosonized
description  that allows exact treatment of interactions and hence nonperturbative
treatment of plasmons. Precisely, the
plasmon dynamics is described by the action
\begin{equation}
S=\int_{-1}^{1}dp\int_0^\beta d\tau \int_{-L/2}^{L/2} \frac{dx}{2K_p}\left\{
\frac{1}{u_p}|\partial_{\tau} \Theta_p|^2+u_p|\partial_x\Theta_p|^2\right\},
\label{Sp}
\end{equation}
representing a finite size generalization of the sliding Luttinger liquid
model \cite{SLL}. The relation of the plasmon velocity $u_p$ and the
Luttinger liquid constant $K_p$ with inter- and intrawire interactions has
been calculated  in Ref. \cite{SLL}.
A fermion annihilation operator in the wire $n$, $\hat{\Psi}_n(x)$, is represented as
\begin{equation}
\hat{\Psi}_{n}^{\chi}(x)\sim
\hat{F}_{n}^{\chi}
\exp\left[-i\int_{-1}^{1}dp \phi_{p}^{\chi}(x) e^{-i\pi pn}\right],
\label{Psi-n}
\end{equation}
where $\chi=R,L$ denotes the chirality, $\phi_{p}^{\chi}(x)$ is a chiral bosonic field,
and $\hat{F}_{n}^{\chi}$ is a Klein factor.  The chiral field $\phi_p^{\chi}$ is in turn
expressed through the field $\Theta_p(x)$ and its dual $\Phi_p(x)$,
\begin{equation}
\phi_p^{R,L}(x)=\left(\Theta_p(x)\pm\Phi_p(x)\right)\sqrt{\pi}.
\end{equation}
In the bosonized representation (\ref{Psi-n}),
the correlation function $X(\tau)$  factorizes in the correlator of Klein factors and the
correlator of bosonic exponents that we denote as $X_b(\tau)$.
The time dependence of the Klein factors $F_{n,\chi}(\tau)$ is given
by the ground state energy of the wire $n$ that includes the capacitive interaction between the
wires.  Thus the correlator of the Klein factors is proportional to $e^{-E_c\tau}$, where $E_c$ denotes the
charging energy.
Denoting the correlator of bosonic exponents as
\begin{equation}
X^{\chi\chi'}_b(\tau)=\left\langle T_{\tau}\left(e^{-i\phi_{0\chi}(0,\tau)}
e^{i\phi_{1\chi'}(0,\tau)}e^{-i\phi_{1\chi'}(0,0)}e^{-i\phi_{0\chi}(0,0)}\right)\right\rangle
\label{XBtau}
\end{equation}
we can cast the expression for the current into the form
\begin{equation}
I_1(V)=-\frac{2\pi e}{\hbar}\int_{-\infty}^{\infty}dt \sum_{\chi,\chi'=L,R}
\left[e^{i(\omega+E_c) t}X^{\chi\chi'}_{b-}(\tau=it+0)-
e^{i(\omega-E_c) t}X^{\chi\chi'}_{b+}(\tau=it+0)\right].
\label{IVc}
\end{equation}
Here $\chi,\chi'=R,L$ denote the chirality of corresponding boson field. We assume the size of the tunneling region along the wire much larger than the Fermi wave length. Then the terms with equal chiralities, $\chi=\chi'$, give the major contribution to the current (\ref{IVc}). Leaving only those terms and noticing  that the contributions form the modes with left and right chiralities to the tunneling current are equal, we can write the conductance of a single junction as
\begin{equation}
\sigma_1(V)=\frac{e^2}{h}\left[\frac{d}{d\omega}X_b(\omega)\bigg|_{\omega=-E_c}-
\frac{d}{d\omega}X_b(\omega)\bigg|_{\omega=E_c}\right],
\label{cond1}
\end{equation}
Substituting the explicit form of the correlator of free bosonic fields, assuming a symmetric form of the dispersion $u(p)=u(-p)$, we finally cast the correlator
(\ref{XBtau}) to the form ($0\leq\tau\leq\beta$)
\begin{equation}
X_{b+}(\tau)=\exp\left[-\left\langle\kappa_p
S_p(\tau,a)\right\rangle\right],
\label{XB+}
\end{equation}
where $\kappa_p=K_p+1/K_p$,
\begin{eqnarray}
\nonumber &&
S_p(\tau,a)=\sum_{m=0}^{\infty}\left\{
\ln\sinh\left[\frac{\pi u_p}{2L}\left((m+1)\beta-\tau\right)\right]
+\ln\sinh\left[\frac{\pi u_p}{2L}(m\beta+\tau)\right]
\right. \\
&&
\left.
-\ln\sinh\left[\frac{\pi u_p}{2L}\left((m+1)\beta-a\right)\right]
-\ln\sinh\left[\frac{\pi u_p}{2L}(m\beta+a)\right]
\right\},
\label{S_p}
\end{eqnarray}
and the average over the plasmon wave vector $p$ is defined by
\begin{equation}
\langle \cdot \rangle_p\equiv \int_0^{1}\cdot \left(1-2\cos(\pi p)+\cos(2\pi p)\right)dp.
\end{equation}
For $-\beta<\tau<0$ the correlator assumes the form
\begin{equation}
X_{b-}(\tau)=\exp\left[-\left\langle\kappa_p
S_p(-\tau,-a)\right\rangle\right].
\label{XB-}
\end{equation}
In Eqs. (\ref{XB+}), (\ref{S_p}), (\ref{XB-}), the zero temperature result is given by the terms in
(\ref{S_p}) with $m=0$, whereas the terms with $m>0$ give the finite temperature correction.

Further we assume the coupling constant  to be $p$-independent,
$\kappa_p=\kappa$ and use the simplified dispersion law (\ref{u_p}).
Note, that in approximate evaluations it is much more important to keep the
$p$-dependence of the velocity $u_p$ that reflects the formation of plasmon bands
than the $p$-dependence of the coupling constant $\kappa_p$. To the lowest order in
$p$-dependent terms, the latter just leads to the averaging of the single Luttinger liquid
result over the coupling constant.

Expanding the logarithmic functions in (\ref{S_p}) we can rewrite
(\ref{XB+}), (\ref{XB-}) in the form
\begin{equation}
X_{b\pm}=I^0_{\pm}\prod_{m=1}^{\infty}Z_m(\tau),
\label{X_b:Z}
\end{equation}
where
\begin{equation}
Z_m(\tau)=\exp\left\{\kappa\sum_{n=1}^{\infty}\left\langle e^{-\frac{\pi u_p n}{L}(m\beta+\tau)}
+e^{-\frac{\pi u_p n}{L}(m\beta-\tau)}\right\rangle_p^{\rm reg}\right\}.
\label{Zm}
\end{equation}
Here and everywhere further ``${\rm reg}$'' denotes the short time regularization, which for a function
$f(\tau)$ is defined as follows
\begin{equation}
f(\tau)^{\rm reg}=f(\tau)-f(a)
\label{reg}
\end{equation}
$a$ being a short-time cutoff. The function $I_{\pm}(\tau)$ is defined as
\begin{equation}
I^0_{\pm}(\tau)=\exp\left\{\mp\kappa\frac{\pi\langle u_p\rangle_p}{2L}\tau+\kappa\sum_{n=1}^{\infty}\left\langle e^{\mp\frac{\pi u_p n}{L}\tau}\right\rangle_p^{\rm reg}\right\}.
\label{I0}
\end{equation}

\section{Resistance at low temperatures}
\label{sec-lowT}

The low temperature transport regime is determined by the condition that the temperature is less than
the distance between the neighbor plasmon bands, $T< \pi v_0/L$.
In terms of (\ref{X_b:Z}), (\ref{Zm}) and (\ref{I0}) the current through a break (\ref{IVc}) can be represented as
a convolution in frequency space
\begin{equation}
I_1(V)=-\frac{2\pi e}{\hbar}|t_{01}|^2\int_{-\infty}^{\infty}d\omega
\left[I^0_{-}(\omega+E_c-eV)-I^0_{+}(\omega-E_c-eV)\right]Z(-\omega),
\label{I1-om-Z}
\end{equation}
where $I^0_{\pm}(\omega)$  denote the real time Fourier transform of the function
$I^0(t)$, and $Z(\omega)$ is the Fourier transform of $Z(t)=\prod_{m=1}^{\infty}Z_m(t)$.
One can consider (\ref{I1-om-Z}) as a kind of Fermi golden rule formula by rewriting it as
\begin{equation}
I_1(V)=e\left(\Gamma_{+}(V)-\Gamma_{-}(V)\right)
\label{I1-FGR}
\end{equation}
with obvious identifications for the transition rates
\begin{equation}
\Gamma_{\pm}=\frac{2\pi}{\hbar}|t_{01}|^2\int_{-\infty}^{\infty}d\omega
I^0_{\pm}(\omega\mp E_c-eV) Z(-\omega).
\label{Gamma-plasmon}
\end{equation}
Then the quantity $I^0_{\pm}$ describes the tunneling density
of states at zero temperature, while the factor $Z(\omega)$ describes
the density of thermally activated {\it plasmon complexes} that play the role of  bosonic
agents assisting the hop.

Expressions (\ref{Zm}), (\ref{I0}) contain the averaging over the plasmon wave vector $p$ that
mathematically describes the influence of plasmon bands on transport. After the analytical continuation
to real times $t=i\tau$, the basic average to be used in subsequent calculations reads
\begin{equation}
\langle e^{-bu_p}\rangle_p=e^{-bv_0}\left[I_0(b v_1)+(1-\frac{1}{bv_1})
I_1(bv_1)\right],
\label{av-dp-I}
\end{equation}
where $b=(m\beta \pm it)\pi/L$, and $I_{\nu}(z)$ denotes the Bessel function of complex argument.
The time $t$ in (\ref{av-dp-I}) is limited by the hopping time over the break $t_h$. In what follows
we concentrate on the regime of long hopping times $t_h$ that is defined by the condition
 $1/t_h \ll \frac{v_1}{L}$.
In  that regime the charge density fluctuations from spatially separated regions can reach the break during the hopping time. Thus, the full plasmon spectrum can participate in the creation
of a bosonic excitation that assists the hopping process. Mathematically, this fact is reflected  in using
the approximation of Bessel functions at large value of the argument in the form
\begin{eqnarray}
\langle e^{-\frac{\pi u_p}{L}(m\beta\pm it)}\rangle_p\approx
\sqrt{\frac{2L}{\pi^2|v_1|(m\beta\pm it)}}e^{\left[-\frac{\pi}{L}w(m\beta\pm it)\right]},
\label{av-p} &&
\end{eqnarray}
where $w=v_0-v_1$.
Despite being obtained for $v_1<v_0$, (\ref{av-p}) is essentially
nonperturbative in $v_1$, which reflects a highly nonlinear influence of plasmons on the
thermally activated transport. We note in passing that the consideration of the long hopping time regime is in accord with the identification of the break as a junction with an exponentially high resistance.

\subsection{Zero temperature tunneling density of states}

Performing the analytical
continuation to real times in (\ref{I0}) and using the large time approximation (\ref{av-p}) for $m=0$,
we obtain $I^0_{\pm}(t)$ as an expansion
\begin{equation}
I^0_{\pm}(t)=e^{\mp i(E_c+\kappa\frac{\pi w}{2L}) t}\sum_{l=0}^{\infty}\frac{\kappa^l}{l!}\left(\frac{2L}{\pm i\pi^2|v_1|t}\right)^{l/2}
\left(\sum_{n=1}^{\infty}\frac{1}{n^{3/2}}e^{\mp in\frac{\pi w}{L}t}\right)^l.
\label{I0_expansion}
\end{equation}
In performing the Fourier transformation in (\ref{I0_expansion}) term by term, one meets the
following typical Fourier transform
\begin{equation}
\int_{-\infty}^{\infty}dt\frac{e^{i\omega t}}{[-i(t\pm io)]^{n/2}}=
2\sin\left(\frac{\pi n}{2}\right)\Gamma\left(1-\frac{n}{2}\right)
\theta(\mp \omega)|\omega|^{n/2-1}.
\label{I0-FT}
\end{equation}
It follows from (\ref{I0-FT}) that each term in the Fourier transform of (\ref{I0_expansion}) is proportional to a step function of the type
$\theta\left(\omega-E_a-\frac{\pi k w}{L}\right)$, the integer $k$ being given a sum of $l$ numbers
of $n_1, ..., n_l$ from the sum over $n$ in (\ref{I0_expansion}). Since $\omega=eV$, only the terms with lowest
$l$ and $n$ are important in the linear transport regime at small $\omega$. Leaving only the terms with
$l=0,1$ and $n=1$, we obtain
\begin{equation}
I^0_{\pm}(\omega)=\delta(\omega\mp E_a)
+2\kappa\sqrt{\frac{2L}{\pi^2|v_1|}}\Gamma\left(\frac{1}{2}\right)
\theta\left(\pm\omega-E_a-\frac{\pi}{L}w\right)\left|\pm\omega-E_a-\frac{\pi}{L}w
\right|^{-1/2},
\label{I0-1}
\end{equation}
where we introduced the addition energy $E_a=E_c+\frac{\pi\kappa}{2L}\langle{u_p}\rangle_p$.
Appearance of a continuous tunneling density of states at finite bias $\omega=eV>E_a+\frac{\pi}{L}w$
indicates a strongly correlated nature of the system under consideration. It is due to the highly nonlinear
interaction of electrons and charge density waves. This feature, which is not relevant
for zero temperature linear response because of the finite energy gap to the continuous part of the spectrum, will determine the thermally activated transport that is
described below. For the zero temperature tunneling density of states  we restrict ourselves with the zero approximation leaving only the delta-function term in (\ref{I0-1}).

\subsection{Influence of thermally activated plasmons.}

As it has been pointed out above, the influence of
thermally activated plasmons is contained in the factor $Z(\omega)$ (\ref{I1-om-Z}) that is
given by the Fourier transform of
\begin{equation}
Z(t)=\exp\left[-\left\langle\kappa_p\sum_{\sigma=\pm 1}\sum_{m=1}^{\infty}
\ln\left(1-e^{-\frac{\pi u_p}{L}(m\beta+i\sigma t)}\right)\right\rangle_p
\right].
\label{Z(t)wd}
\end{equation}
At low temperatures, $T\ll\frac{\pi w}{L}$, the major contribution to $Z(t)$ is given by the
term with $m=1$.
Leaving only that term, expanding the logarithms in (\ref{Z(t)wd}) and using (\ref{av-p})  we obtain
\begin{equation}
Z(\omega)\approx\sum_{n,l=0}^{\infty}\frac{2\kappa^{l+n}}{n!l!}
\left(\frac{2L}{\pi^2|v_1|}\right)^{\frac{n+l}{2}}
e^{-\beta\left(\frac{\pi w}{L}(n+l)+|\omega-\epsilon_{ln}|\right)}
\left|\omega-\epsilon_{ln}\right|^{\frac{\nu}{2}-1}
\frac{\sin\left(\frac{\pi\nu}{2}\right)\Gamma\left(1-\frac{\nu}{2}\right)}{\left(2\beta
\right)^{\frac{l+n-\nu}{2}}},
\label{Z1}
\end{equation}
where $\epsilon_{ln}(\omega)=\frac{\pi w}{L}(l-n)$, and $\nu=l,n$ for
$\omega-\epsilon_{ln}\gtrless 0$ respectively.
Each term in (\ref{Z1}) describes a thermal excitation of a multiparticle plasmon complex with
a continuous density of states.
The plasmon complexes
described mathematically by (\ref{Z1}) form the bath of bosonic excitations that facilitate
transport over the break.
Higher values of $m$ in (\ref{Z(t)wd}) would correspond to excitations involving progressively
more plasmon modes. The leading terms in (\ref{Z1}) at low temperatures are given by
$l,n=\overline{0,1}$. Those terms result in the leading low-temperature contribution to $\sigma_1$ in the form
\begin{equation}
\sigma_1\approx\frac{e^2}{h}\frac{\kappa^2L\Gamma\left(\frac{1}{2}\right)}{
\sqrt{2}\pi^2|v_1|T}\left(|\beta E_a|^{-\frac{3}{2}}
+2|\beta E_a|^{-\frac{1}{2}}\right)
e^{-\beta\left(E_a+\frac{2\pi}{L}w\right)},
\label{si1-approx}
\end{equation}
from which we identify
\begin{equation}
f(E_a,T)=\beta\left(|E_a|+\frac{2\pi}{L}w\right)-
\ln\left(\frac{1}{2}|\beta E_a|^{-3/2}+|\beta E_a|^{-1/2}\right).
\label{f(E,T)-lowTd1}
\end{equation}

\subsection{Resistance of a long array}

Further evaluation of the resistance proceeds according to Eqs. (\ref{R1}), (\ref{area}), (\ref{R_l}) of
Sec. \ref{sec-RR}.
The upper limit of integration in (\ref{area})
\begin{equation}
u-f(E_0,T)=0
\label{upperlimit}
\end{equation}
relates the value of $u$ to the corresponding value
of the addition energy $E_0$. Under the assumption $\beta E_0\gg\ln(\beta E_0)$ that will be selfconsistently confirmed by the solution below,  the integration in (\ref{area}) results in
\begin{equation}
A(E_0,T)=d \left\{\frac{E_0^2}{2T}+\frac{T}{2}\ln\left(\frac{2E_0}{T}+1\right)+
\frac{E_0}{2}\right\}.
\label{A(E_0)}
\end{equation}
The saddle point value of $u$ in (\ref{R_l}) relates to the saddle point value of the energy $E_0$ which
is obtained as
\begin{equation}
E_0\big|_{SP}=\frac{1}{gd}.
\label{E_0SP}
\end{equation}
Therefore, the assumption above is justified for low temperatures, when $1/(T g d)\gg 1$.
Further evaluation of (\ref{R_l}) leads to
the following result for the resistance of the array
\begin{equation}
R\approx l_y\frac{h}{e^2}
\frac{\pi^2|v_1|  T^{2}}{\sqrt{2}\kappa^2L\Gamma\left(\frac{1}{2}\right)} \exp\left[\left(\frac{1}{2gdT}+\frac{\pi w}{LT}\right)\right].
\label{R(T)}
\end{equation}
For comparison, as it will be shown below in Sec. \ref{sec-phonon},
in the case of phonon-assisted transport in the same model the preexponential
factor in (\ref{R(T)}) goes like $T^{1/2}$.

\subsection{Low temperatures, short array.}

In the case of a short array, the optimal break cannot be found in the array. In that
case the highest resistance amongst all those present in the array determines the total resistance.
That highest resistance is defined by the condition $l_y\rho(u_f)\sim 1$, where $l_y$ denotes the
length of the array, and $\rho(u)=e^{-gA(u)}$ denotes the probability density for the logarithmic
resistance $u=\log(R/R_0)$ \cite{RR}. The value $u_f$ is related to the length of the array by the
condition
\begin{equation}
A(u_f)=\frac{1}{g}\ln{l_y}.
\label{Af}
\end{equation}
Furthermore, the condition for the upper limit of integration $u_f=f(E_f,T)$
in (\ref{area}) determines the characteristic value of the addition energy $E_f$.
The area $A(u_f)$ is then given by (\ref{A(E_0)}) with $E_0$ replaced by $E_f$,
that together with (\ref{Af}) provides the relation between the length of the array $l_y$ and the
characteristic energy $E_f$.  For $E_f/T\gg 1$, (\ref{A(E_0)}) can be simplified to
\begin{equation}
A\approx d \frac{E_f^2}{2T}=\frac{1}{g}\ln l_y,
\label{Ef--Ly}
\end{equation}
where we used (\ref{Af}) is the last equality.
Finally, the resistance of the array is given by $R=R_0e^{u_f}$ with $u_f=f(E_f,T)$. Expressing $E_f$ as
a function of $l_y$ with the help of (\ref{Ef--Ly}), we obtain the resistance of the short array in the form
\begin{equation}
R(l_y,T)\approx \frac{R_0\exp\left[\frac{2\pi w}{LT}+\left(\frac{2\ln l_y}{agT}\right)^{1/2}\right]}{\left( \frac{agT}{2\ln l_y}\right)^{1/4}+\frac{1}{2}\left( \frac{agT}{2\ln l_y}\right)^{3/4}}.
\label{R(L_y,T)}
\end{equation}
Expression (\ref{R(L_y,T)}) determines the temperature dependence of the short array and is valid for the length of the array $l_y<e^{\frac{1}{2gaT}}$. In the opposite regime,  $l_y>e^{\frac{1}{2gaT}}$, the resistance is given by Eq. (\ref{R(T)}).

\subsection{Plasmon assisted resistance for short hopping time}

The integral over the time in (\ref{IVc}) should actually be cut off by the hopping time $t_h$.
In the regime of  very low temperatures and short hopping time over the break
$T\ll \frac{v_1}{L}$, $\frac{L}{v_1} \gg t_h \gg \frac{L}{v_0} $,
the large argument expansion (\ref{av-p}) applies for $m\neq 0$ while a short time expansion should be used in calculating $I^0_{\pm}(t)$. We obtain $I^0_{\pm}(t)= 1/2+ O\left(\frac{\pi v_1 t}{L}\right)$ from which it follows $I^0_{\pm}(\omega)=\frac{1}{2}\delta(\omega)$. For $Z(t)$, leaving only the factor with $m=1$, we get
\begin{equation}
Z(t)\approx 1+2\kappa\sqrt{\frac{2L T}{\pi^2 v_1}}e^{-\beta\frac{\pi w}{L}}\cos\left(\frac{\pi w}{L}t
\right)+O\left(e^{-2\beta\frac{\pi w}{L}}\right),
\label{Z(t)-short}
\end{equation}
form which it follows
\begin{equation}
Z(\omega)\approx 2\pi\delta(\omega)+2\pi\kappa\sqrt{\frac{2L T}{\pi^2 v_1}}e^{-\beta\frac{\pi w}{L}}
\left[\delta\left(\omega-\frac{\pi w}{L}\right)+\delta\left(\omega+\frac{\pi w}{L}\right)\right].
\label{Z(omega)-short}
\end{equation}
Eq. (\ref{Z(omega)-short}) shows that the bosonic spectrum remains discrete, corresponding to the spectrum of a single wire. Physically it means that the
plasmons from the distant wires cannot reach the break during the hopping time, and the existence of
plasmon bands plays no role for the electronic transport. In that regime, the influence of interactions reduces to the charging energy of a finite wire, which is the main assumption in the models of granular metals with high intergrain conductance \cite{AGK,Efetov}.
Finally we note that the condition of short hopping time may contradict the assumption of an exponentially high resistance of a break.

\section{Resistance at high temperatures}
\label{sec-highT}

At high temperatures, when $T\gg\frac{\pi w}{L}$, the temperature broadening exceeds the
interlevel separation in a single wire. To analyze the temperature dependence of resistance in that regime, it is convenient to rearrange the sum over $m$ in (\ref{S_p}) with help of the Poisson summation formula
\begin{eqnarray}
\nonumber &&
S\equiv\sum_{\sigma=\pm 1}\sum_{m=1}^{\infty}
\ln\left(1- e^{-\frac{\pi u_p}{L}(m\beta+i\sigma t)}\right)
=\sum_{m=-\infty}^{\infty}
\ln\left(\frac{\sinh\left[\frac{\pi u_p}{2L}\left(m\beta+\tau\right)
\right]}{\sinh\left[\frac{\pi u_p}{2L}\left(m\beta+a\right)\right]}\right)\\
 &&
=\int_{-\infty}^{\infty}dx\ln\left(\frac{\sinh\left[\frac{\pi u_p}{2L}\left(\beta x+\tau\right)
\right]}{\sinh\left[\frac{\pi u_p}{2L}\left(\beta x+a\right)\right]}\right)+
2\sum_{k=1}^\infty \int_{-\infty}^{\infty}dx\cos(2\pi kx)
\ln\left(\frac{\sinh\left[\frac{\pi u_p}{2L}\left(\beta x+\tau\right)
\right]}{\sinh\left[\frac{\pi u_p}{2L}\left(\beta x+a\right)\right]}\right).
\label{sumP1}
\end{eqnarray}
The last line of (\ref{sumP1}) has been obtained using the Poisson summation formula.
After rescaling the integration variable in the integral over $x$ as
$y=\frac{\pi u_p}{2L}\beta x$, we obtain
\begin{equation}
S=\frac{2L}{\pi\beta u_p}\int_{-\infty}^{\infty}dy\ln
\left(\frac{\sinh\left[y+\frac{\pi u_p}{2L}\tau\right]}{\sinh
\left[y+\frac{\pi u_p}{2L}a\right]}\right)
\left[1+2\sum_{k=1}^{\infty}\cos\left(\frac{4kL}{\beta u_p}y\right)\right].
\label{rescaledS}
\end{equation}
Integrating by parts in each term with a given $k$, we finally  obtain
\begin{eqnarray}
&&
S=\sum_{k=1}^{\infty}S_k,
\label{S-Sk} \\
&&
S_k=\frac{1}{k}\left[\cos\left(2\pi k\frac{a}{\beta}\right)
-\cos\left(2\pi k\frac{\tau}{\beta}\right)\right]
\coth\left(\frac{2\pi k L}{\beta u_p}\right).
\label{Skfin}
\end{eqnarray}
Let us  rearrange the sum for $S$ in form of a perturbative
expansion around the result for a Luttinger liquid of infinite length.
To this end we note, that if we replace $\coth\left(\frac{2\pi k L}{\beta u_p}\right)= 1$
for all $k$, then the sum over $k$ reads
\begin{equation}
\sum_{k=1}^{\infty}S_k\approx \ln\left|\frac{\sin(\pi \tau/\beta)}{\sin(\pi a/\beta)}
\right|,
\label{0-approx}
\end{equation}
which is the result for the bosonic correlator in the infinite length Luttinger liquid.
Therefore, the expression for $S$ can be written in the form
\begin{equation}
S=\log\left|\frac{\sin(\pi\tau/\beta)}{\sin(\pi a/\beta)}\right|+
\sum_{k=1}^{\infty}\frac{1}{k}\left[\cos\left(2\pi k\frac{a}{\beta}\right)
-\cos\left(2\pi k\frac{\tau}{\beta}\right)\right]
\left[\coth\left(\frac{2\pi k L}{\beta u_p}\right)-1\right].
\label{inftL-expan}
\end{equation}
Here the first term is the infinite length result, and the sum over $k$ denotes
the corrections due to the finite length of the wire $L$. For large argument of
$\coth$, $\left(\frac{2\pi k L}{\beta u_p}\right)\gg1$, the $k$-correction decays
as $\frac{1}{k}\exp\left(-\frac{4\pi k L}{\beta u_p}\right)$.
Further calculation is performed for the zero order term in  $1/L$ expansion.
Evaluating the Fourier transform of $X_b(\tau)=e^{-\left\langle\kappa_p S(\tau)\right\rangle_p}$ and performing the analytical continuation to real frequencies, we obtain
\begin{equation}
X_b(\omega)\approx -\frac{(2\pi a)^{\kappa}}{\pi\beta^{\kappa-1}}
\sin(\pi\kappa/2)\Gamma(1-\kappa)
\frac{\Gamma\left(\kappa/2+i\frac{\beta\omega}{2\pi}\right)}{\Gamma\left(1-\kappa/2+
i\frac{\beta\omega}{2\pi}\right)}.
\label{Xb_T}
\end{equation}
The conductance of a single junction is then calculated by (\ref{cond1}). Substituting $\omega=-E_c$ in
(\ref{Xb_T}) one can see that the linear current through the junction and hence its conductance does
not fluctuate exponentially strongly with the charging energy $E_c$.
Therefore, the resistance of the array is not determined by the resistance of a single break, it is rather given by the average over the resistances of all junctions. This result also suggests that at high enough temperatures the exponential suppression factor $e^{-E_c/T}$ that is typical for the activated transport
is compensated and even overridden by the exponential increase in the number of particle hole pairs that can provide the energy $E_c$ to hop over the break.

The fact that the leading contribution to the conductance at high temperatures is given by the result for infinitely long wires implies that the coherence of the single particle motion is broken already by a single hop between the two neighbor wires. The latter justifies the averaging over the all junctions in calculation of the resistance. In the result, the expression for the resistance can
be written as a Drude formula $\sigma=e^2 N\tau_f/m_*$ ($N$ is the electron concentration and $m_*$ is the effective mass)
with the interaction and temperature dependent mean free time
$\tau_f(T)$. The expression for the mean free time
can be organized as an expansion in powers of the small parameter
$e^{-\frac{4\pi LT}{w}}$.

At a typical charging energy $\bar{E}_c\gg 2\pi T$, the derivative  $dX_b/d{\omega}$ for $\omega=-E_c$ in (\ref{cond1}) can be approximated as
\begin{equation}
\frac{dX_b}{d\omega}\bigg|_{\omega=-E_c}\approx -2a^{\kappa}\sin\left(\frac{\pi\kappa}{2}\right)
\Gamma(2-\kappa)E_c^{\kappa-2}.
\label{E_kappa-2}
\end{equation}
Thus, in that regime the
leading term in the expression for the mean free time is temperature independent,
$\tau_f\propto 1/\langle E_c^{2-\kappa}\rangle$. For comparison, phonon assisted transport in
that temperature regime still has a thermally activated character with the preexponential
factor $T^{-1/2}$.

At temperatures even larger than the typical charging energy,
$\bar{E}_c<T$, the expansion of $\frac{dX_b}{d\omega}\bigg|_{\omega=-E_c}$  in small parameter
$\frac{\beta E_c}{2\pi}$  results in the following expression for the conductance of the single junction
\begin{equation}
\sigma_1=\frac{e^2}{h}\frac{(2\pi a)^{\kappa}}{2\pi^3}\sin\left(\frac{\pi\kappa}{2}\right)
\Gamma(1-\kappa)V'(0)T^{\kappa-3}E_c,
\label{sigma-highT}
\end{equation}
where $V'(0)$ is the derivative $dV(x)/dx|_{x=0}$ of the function
\begin{equation}
V(x)=\frac{\Gamma(\kappa/2+x)}{\Gamma(1-\kappa/2+x)}\left[\psi(\kappa/2+x)-\psi(1-\kappa/2+x)
\right].
\end{equation}
Subsequent averaging of the resistances of all junctions over the random charging energy $E_c$ leads to
the Drude expression for the resistance, where $\tau_f$ exhibits the power-low
temperature dependence typical for transport across  a sliding Luttinger liquid,
$\tau_f\propto T^{3-\kappa}\langle E_c^{-1}\rangle$ \cite{SLL}. Therefore, at high temperatures,
the mean free time is determined by the Luttinger liquid interaction parameter $\kappa$.
The resistance for phonon assisted transport in that regime is given by a Drude formula with
the logarithmic temperature dependence of the mean free time $\tau_f\propto\ln T$.

\section{Phonon assisted transport }
\label{sec-phonon}

In this section we calculate the phonon assisted thermally activated transport in the system
under consideration. For simplicity, we consider the case of noninteracting electrons, taking into account only the charging energy by the transfer of electrons between the wires. Then the
single particle energy spectrum of each wire near the Fermi energy consists of equidistant energy
levels with the interlevel distance $\pi v_0/L$, $v_0$ being  the Fermi velocity. Furthermore, we assume the phonons to have a featureless density of states with the spectral density $D_{\rm ph}=\omega_D/(2\pi v_s)$, where $\omega_D$ is the Debye frequency and $v_s$ is the sound velocity. The phonon assisted transition rate between the wires $0$ and $1$ that form the break can be calculated as \cite{Efros}
\begin{equation}
\Gamma_{01}=|t_{01}|^2\gamma_{\rm e-ph}\sum_{n,l}f(\epsilon_n)[1-f(\epsilon_l-eV)]N_B(E_c+\epsilon_l-
\epsilon_n-eV),
\label{Gamma-ph}
\end{equation}
where $\epsilon_n=\pi v_0 n/L$ is the single particle energy with respect to the Fermi level, $f(\epsilon)$ denotes  the Fermi distribution and $N_B(\omega)$ denotes the Bose distribution, and $\gamma_{\rm e-ph}$ is the electron-phonon coupling constant. The linear conductance of the break is given by
\begin{equation}
\sigma_{\rm break}=\frac{e^2}{T}\Gamma_{01}.
\label{sigma_break}
\end{equation}

\subsection{Low temperature regime}

Calculating the transition rate according to (\ref{Gamma-ph}) we obtain
\begin{equation}
\Gamma_{01}=|t_{01}|^2\gamma_{\rm e-ph}e^{-\beta E_c}\left\{\frac{1}{1-e^{-\beta\frac{\pi v_0}{L}}}+
\frac{\beta\frac{\pi v_0}{L} e^{-\beta\frac{\pi v_0}{L}}}{\left(1-e^{-\beta\frac{\pi v_0}{L}}\right)^2}
\right\}
\label{Gamma01}
\end{equation}
Further calculation follows the procedure described in Section \ref{sec-RR} and leads to the
phonon assisted thermally activated resistance in the form
\begin{equation}
R_{\rm phonon}\approx l_y\sqrt{\frac{2\pi T}{g d}}e^{\frac{1}{g d T}}.
\label{R-phonon}
\end{equation}

\subsection{High temperature regime}

In the high temperature regime, $T\gg \frac{\pi v_0}{L}, \  T>E_c$, we replace in (\ref{Gamma-ph})
$f(\epsilon_n)\approx 1-f(\epsilon_l-eV)\approx 1/2$, and replace the summation over $n,l$ by integration
\begin{eqnarray}
\nonumber &&
\Gamma_{01}\approx  \frac{|t_{01}|^2\gamma_{\rm e-ph}}{4}\int_0^\infty \frac{dx}{e^{\beta E_c}
e^{\beta\frac{\pi v_0}{L}x}-1} \\
&&
=-\frac{|t_{01}|^2 LT\gamma_{\rm e-ph}}{4\pi v_0}\ln\left|1-e^{-\beta E_c}\right|
\approx -\frac{|t_{01}|^2 LT\gamma_{\rm e-ph}}{4\pi v_0}\ln\left|\beta E_c\right|.
\label{Gamma-pn-highT}
\end{eqnarray}
Substituting that result into (\ref{sigma_break}), we obtain
\begin{equation}
\sigma_{01}\approx\frac{e^2|t_{01}|^2 L\gamma_{\rm e-ph}}{4\pi v_0}\ln\left|\beta E_c\right|.
\label{sigma-pn-highT}
\end{equation}
As argued above, the resistance of the whole array is given by the average over the resistances of all
junctions over the random charging energy $E_c$ and the random hopping matrix elements $t_{ij}$,
which results in
\begin{equation}
R_{\rm phonon}=l_y\frac{4\pi v_0}{e^2\gamma_{\rm e-ph} l_y}\left\langle|t_{ij}|^2\ln\left|\frac{T}{E_c}
\right|\right\rangle.
\label{R-phonon-heighT}
\end{equation}
Therefore, at temperatures exceeding the typical addition energy, the resistance of phonon assisted transport has a logarithmic dependence of temperature.

\section{Conclusion}
\label{sec-concl}

In this paper we calculated temperature dependence of resistance for plasmon assisted thermally
activated transport in a particular model of a one-dimensional granular array. The considered model is  relevant for experiments on transport across a highly anisotropic two-dimensional systems that can be
represented as an array of parallel one-dimensional wires \cite{Mani,deHeer,Himpsel,Fogler}.

We showed the existence of the regime, where the localization length of the plasmon excitations exceeds by far the single particle localization length. The latter allows plasmons to traverse the whole array and thus form an external bath of bosonic excitations necessary to suppress single particle Anderson localization \cite{Malinin}. Further  peculiar property of the model lies in its closeness to the sliding Luttinger liquid model \cite{SLL} as the description of interactions concerns.
 On one hand this property allows to use a powerfull bosonization technique and to arrive at conclusions about the temperature dependence of the {\it preexponential} factor in the thermally activated resistance (\ref{R(T)}). The calculation also gives an insight in the formation of multiplasmon complexes that create a bosonic bath with continuous density of states (\ref{Z1}). Moreover, the increase of the number of particle-hole excitation with temperature leads to the suppression of the thermally activated behavior at high temperatures, the resistance being given by the Drude formula with the mean free time that depends on temperature as a power law. The power low dependence of the Drude mean free time in the hight temperature regime is in general agreement with findings of the recent work \cite{Gornyi}. At this point one might speculate about the analogy between that change of the character of transport and  the recently reported finite temperature phase transition form the Anderson localized to a delocalized phase in interacting disordered systems \cite{Aleiner,Mirlin}.

On the other hand, our results show that the capacitive model used for the description of transport in granular arrays \cite{AGK,Efetov,Fogler03} is inapplicable for that particular kind of array. This property again stems from the closeness of our model to the sliding Luttinger liquid one, where the relaxation time of charge density excitations is very long. The chosen specific model of a one dimensional array enabled us to show that thermally activated resistance has qualitatively different temperature dependence
for plasmon assisted transport as compared to phonon assisted transport.

Despite the specific quasi one-dimensional geometry of the grains in this model,
the present investigation is believed to be of general importance for granular arrays with
delocalized or weakly localized plasmons.

\begin{acknowledgments}
The author is grateful to M. Raikh, who initiated this work, for numerous illuminating
discussions. The author appreciates fruitful discussions with D. Pfannkuche and
valuable comments of I. Gornyi and S. Kettemann. Financial support by Deutsche
Forschungsgemeinschaft through SFB 508 is gratefully acknowledged.
\end{acknowledgments}


\begin{thebibliography}{20}
\bibitem{Berezinskii} V. L. Berezinskii, Zh. Eksp. Teor. Fiz. {\bf 65}, 1251 (1973)
[Sov. Phys. JETP {\bf 38}, 620 (1973)].
\bibitem{Efros} B. I. Shklovskii and A. L. Efros,
{\it Electronic Properties of Doped Semiconductors} (Springer-Verlag, Berlin, 1984).
\bibitem{Tigran} T. V. Shahbazyan and M. E. Raikh, Phys. Rev. B {\bf 53}, 7299 (1996).
\bibitem{Gornyi}I. V. Gornyi, A. D. Mirlin, D. G. Polyakov, arXiv:cond-mat/0407305.
\bibitem{Malinin} S. V. Malinin, T. Nattermann, B. Rosenow, arXiv:cond-mat/0403651.
\bibitem{Fogler03} M. M. Fogler, S. Teber and B. I. Shklovskii, Phys. Rev. B 69, 035413 (2004);
T. Nattermann, T. Giamarchi and P. Le Doussal, Phys. Rev. Lett. {\bf 91}, 056603 (2003).
\bibitem{JvD} J. von Delft \and H. Schoeller, Ann. Phys. (Leipzig) {\bf 7}, 225, (1998).
\bibitem{Haldane} F. D. M. Haldane, J. Phys. C {\bf 14}, 2585 (1981).
\bibitem{Kane-Fisher} C. L. Kane \and  M. P. A. Fisher, Phys. Rev. Lett. {\bf 68}, 1220 (1992).
\bibitem{Anderson} P. W. Anderson, Phys. Rev. Lett. {\bf 18}, 1049 (1967).
\bibitem{Mani}R. G. Mani and K. v. Klitzing, \prb, {\bf 46}, R9877 (1992).
\bibitem{deHeer}W. A. de Heer {\it et al.},  Science, {\bf 268}, 845 (1995).
\bibitem{Himpsel}F. J. Himpsel {\it et al.},  J. Phys.: Condens. Matter {\bf 13},
   11097 (2001).
\bibitem{Fogler}For a review, see M. M. Fogler in
{\it High Magnetic Fields: Applications in Condensed Matter Physics and
   Spectroscopy}, (Springer-Verlag, Berlin, 2002).
\bibitem{AGK}A. Altland, L. I. Glazman and A. Kamenev, Phys. Rev. Lett. {\bf 92},
026801 (2004).
\bibitem{Efetov}K. B. Efetov and A. Tschersich, Phys. Rev. B {\bf 67}, 174205 (2003);
G. G\"oppert and H. Grabert, The European Physical Journal B {\bf 16}, 687 (2000);
I. S. Beloborodov, K. B. Efetov, A. Altland, F. W. J. Hekking, Phys. Rev. B {\bf 63},
115109 (2001).
\bibitem{Gramada-Raikh} A. Gramada \and M. E. Raikh, Phys. Rev. B {\bf 55}, 7673 (1997).
\bibitem{RR}M.E. Raikh and I.M. Ruzin, Sov. Phys. JETP  {\bf 68}, 642  (1989).
\bibitem{SLL}R. Mukhopadhyay, C. L. Kane and T. C. Lubensky, Phys. Rev. B, {\bf 64},
045120 (2001).
\bibitem{Mahan} G. Mahan, {\it Many-Particle Physics} (Plenum Press, New York, 1981).
\bibitem{Aleiner} D.M. Basko, I.L. Aleiner, B.L. Altshuler,  arXiv:cond-mat/0506617 (2005).
\bibitem{Mirlin} I.V. Gornyi, A.D. Mirlin, D.G. Polyakov, arXiv:cond-mat/0506411 (2005)
\end{thebibliography}

\end{document}